\documentstyle[amssymb,epsfig,12pt]{article}

\begin{document}

\begin{titlepage}
\bigskip \begin{flushright}
\end{flushright}


\vspace{1cm}

\begin{center}
{\Large \bf {Zitterbewegung of a Model Universe}}\\
\end{center}
\vspace{2cm}
\begin{center}
R. B. Mann{%
\footnote{rbmann@sciborg.uwaterloo.ca}} \\
Department of Physics \& Astronomy, University of Waterloo, \\
Waterloo, Ontario N2L 3G1, Canada\\
\vspace{1cm}
G.L. Murphy{%
\footnote{email: gmurphy@raex.com}}\\
St. Paul's Episcopal Church, Akron, Ohio, U.S.A.\\
\vspace{1cm}
PACs Numbers: 04.60.+n, 98.80.Dr\\
\today\\
\end{center}
\begin{abstract}
We investigate the quantum evolution of the metric operators for
Bianchi-Type I model universes in the Heisenberg picture in order to
remove the need to consider the wave function of the universe and
interpret its "spin" variables. The calculation is analogous to that
of the Zitterbewegung of the Dirac electron. We consider the
behavior of the metric near the classical singularity, and consider
the curvature there.  Although factor ordering questions preclude
the presentation of an unambiguous result for the curvature
invariants, it does seem that the classical $t^{-4}$ divergence of
the Kretschmann scalar is not removed by quantization.
\end{abstract}
\end{titlepage}\onecolumn


\section{Introduction}

The Hamiltonian treatment of simple cosmological models has provided one
useful approach to quantum cosmology. In the classical case the canonical
equations have solutions which represent solutions of Einstein's field
equations, and the hope is that the corresponding quantum equations will
represent a full and consistent theory of quantized gravitation\cite{ref1}.
The Hamiltonians for homogeneous universes in general relativity occur in
squared form, the simplest example being that for empty Bianchi-Type I
spaces with two degrees of freedom:

\begin{equation}
H^{2}=p_{+}^{2}+p_{-}^{2}  \label{e1}
\end{equation}%
where we take $c$, but not $\hslash $ to be one.

The problem of quantizing a system with such a Hamiltonian already had to be
faced in the early days of relativistic quantum theory. The answer which led
to the greatest progress was that of Dirac, who extracted the square-root of
the corresponding expression for a particle by introducing spinors. With $%
H^{2}=p^{2}+m^{2}$, one writes

\begin{equation}
H={\bf \alpha \cdot p}+\beta m  \label{e2}
\end{equation}
with $\left( {\bf \alpha ,}\beta \right) $ a set of anti-commuting $4\times
4 $ matrices. This led, as is well known, to a natural understanding of
electron spin and related phenomena.

The same type of procedure can be carried out for cosmological models with a
few degrees of freedom. With the Hamiltonian given by (1) we write

\begin{equation}
H=\sigma _{+}p_{+}+\sigma _{-}p_{-}  \label{e3}
\end{equation}
where the $\sigma $'s are $2\times 2$ Pauli Matrices. Plane wave solutions
of the resulting Schr\"{o}dinger equations are then easily found. But this
procedure, while mathematically quite simple, raises interpretative
problems. The most obvious difficulty is that the significance of states
corresponding to ``spin up'' and ``spin down'' for a cosmological model
remains obscure\cite{ref2}.

Since this difficulty has to do with interpretation of the state vector, it
may be helpful to follow a route that focuses instead on the dynamical
variables of the model. Rather than work in the Schr\"{o}dinger picture, as
has usually been done in quantum cosmology, we may deal with the problem in
the Heisenberg picture. This approach provided some novel insights into the
dynamics of the Dirac electron, and may be expected to do the same for a
model universe.

\section{The Zitterbewegung of the Electron}

For the Dirac Hamiltonian (\ref{e2}), the Heisenberg equations of motion
give the velocity of the particle as $dx^{i}/dt=[x^{i},H]/i\hslash =\alpha
^{i}$. Proceeding further, we obtain $i\hslash d\alpha ^{i}/dt=-2\alpha
^{i}H-2p^{i}$ , a differential equation with the solution $\alpha
^{i}=A^{i}\exp (-2iHt/\hslash )+p^{i}H^{-1}$ , where the Ai are constant.
Since $\alpha ^{i}=1$ and $p^{i}H^{-1}\leq 1$, the magnitude of the $A^{i}$
must be of order unity. Another integration yields%
\begin{equation}
x^{i}=(i\hslash A^{i}/2)\exp (-2iHt/\hslash )H^{-2}+p^{i}H^{-1}t+a^{i}
\label{e4}
\end{equation}%
with the $a^{i}$ further constants. The second and third terms in (\ref{e4})
represent the motion of a particle that is expected from classical dynamics,
motion in a straight line with constant velocity ${\bf p}/H$. The first
term, on the other hand, represents a quantum oscillation about that
classical motion with high frequency ($>2m/\hslash $) and an amplitude
comparable with the particle's reduced Compton wavelength. This is the
Zitterbewegung\cite{ref3} or ``trembling motion''. When the Dirac equation
is taken to describe a single particle, the peculiarly quantum motion
provides one way of thinking about the spin of that particle.

\section{Quantum Cosmology in the Heisenberg Picture: The Metric}

In the cosmological problem, the co-ordinates canonical to $p$ are denoted
by $\beta $, and we use a time parameter $\Omega $, which is related to the
conventional cosmic time $t$ by $t=(4\pi R3/H)e^{-3\Omega} $, where $R$ is a
constant. (Thus $t\rightarrow 0$ corresponds to $\Omega \rightarrow \infty $%
.) The diagonal spatial metric is given by%
\begin{eqnarray}
g_{11} &=&R^{2}e^{-2\Omega }\exp [2(\beta _{+}+\sqrt{3}\beta _{-})]
\nonumber \\
g_{22} &=&R^{2}e^{-2\Omega }\exp [2(\beta _{+}-\sqrt{ 3}\beta _{-})]
\label{e5} \\
g_{33} &=&R^{2}e^{-2\Omega }\exp [-4\beta _{+}]  \nonumber
\end{eqnarray}

The classical equations of motion with (\ref{e1}) then give simply $\beta
_{\pm }=(p_{\pm }/H)\Omega $ or%
\begin{equation}
\beta _{+}=\Omega \cos \theta ,\quad \beta _{-}=\Omega \sin \theta
\label{e6}
\end{equation}%
if $\beta (0)=0$, where $\cos \theta =p_{+}/H$ and $\sin \theta =p_{-}/H$.
When these expressions are substituted into (\ref{e5}), we obtain a
representation of the well-known Kasner metric\cite{ref4}.

We now move to the corresponding quantum calculation with the Hamiltonian (%
\ref{e3}). If we use $\sigma _{+}\sigma _{-}+\sigma _{+}\sigma _{-}=0$, $%
\sigma _{+}^{2}=\sigma _{-}^{2}=1$, the Heisenberg equations of motion give
us, in analogy with (\ref{e4}),%
\begin{equation}
\beta _{\pm }=(i\hslash B_{\pm }/2)\exp (-2iH\Omega /\hslash )H^{-2}+p_{\pm
}H^{-1}\Omega +C_{\pm }  \label{e5a}
\end{equation}%
Here the $B$'s and $C$'s are constant $2\times 2$ matrices.

It is tempting to think that the $C$'s can essentially be ignored. A similar
constant of integration representing the initial position of a particle can
often be dropped in elementary mechanics by just defining the initial
position to be at the origin of coordinates. But in quantum theory matters
are different, as we can see by considering the elementary problem of the
motion of a non-relativistic free particle with $H=p^{2}/2m$.

The Heisenberg equations of motion easily give $p=$ constant and $%
q=(p/m)t+q_{0}$ . We might be tempted to ignore the constant $q_{0}$ but it
cannot vanish because the commutation relations between $q$ and $p$ require
that $[q_{0},p]=i\hslash $.

Having to retain the $C$'s as additive constants might not seem to be a
serious problem. But in our case the required exponentiation of the $\beta $%
's in the metric mean that we will have multiplicative constant operators
which do not commute with the $p$'s in the metric. This will complicate the
quantum calculations.

The metric components, obtained by substituting (\ref{e5a}) into (\ref{e5}),
are now operators which carry the time-dependence of the model. They are
\begin{eqnarray}
g_{11} &=&(\exp D_{1})R^{2}\exp \left[ i\hslash (B_{+}+\sqrt{ 3}B_{-})\exp
(-2iH\Omega /\hslash )H^{-1}\right] \exp \left[ 2(\cos \theta + \sqrt{ 3}%
\sin \theta -1)\Omega \right]  \nonumber \\
g_{22} &=&(\exp D_{2})R^{2}\exp \left[ i\hslash (B_{+}-\sqrt{ 3}B_{-})\exp
(-2iH\Omega /\hslash )H^{-1}\right] \exp \left[ 2(\cos \theta - \sqrt{ 3}%
\sin \theta -1)\Omega \right]  \label{e8} \\
g_{33} &=&(\exp D_{3})R^{2}\exp \left[ -2i\hslash B+\exp (-2iH\Omega
/\hslash )H^{-1}\right] \exp \left[ -2\Omega (1+2\cos \theta )\right]
\nonumber
\end{eqnarray}%
where the exponentiated $D$'s are the multiplicative constant operators
noted above.

In terms of the cosmic time $t$ we have

\begin{eqnarray}
g_{11} &=&(\exp D_{1})R^{2}\exp \left[ i\hslash (B_{+}+\sqrt{3}%
B_{-})(Ht/S)^{2iH/3\hslash }H^{-1}\right] (Ht/S)^{2(\cos \theta +\sqrt{3}%
\sin \theta -1)/3}  \nonumber \\
g_{22} &=&(\exp D_{2})R^{2}\exp \left[ i\hslash (B_{+}-\sqrt{3}%
B_{-})(Ht/S)^{2iH/3\hslash }H^{-1}\right] (Ht/S)^{2(\cos \theta -\sqrt{3}%
\sin \theta -1)/3}  \label{e9} \\
g_{33} &=&(\exp D_{3})R^{2}\exp \left[ 2i\hslash B_{+}(Ht/S)^{2iH/3\hslash
}H^{-1}\right] (Ht/S)^{\left( 1+2\cos \theta \right) }  \nonumber
\end{eqnarray}%
where $S=4\pi R^{3}$. The third exponential factors in (\ref{e8}) and (\ref%
{e9}) give the classical Kasner behavior, while the preceding factors
contain the distinctive quantum mechanical behavior. It should be noted,
though, that $\cos \theta $ and $\sin \theta $, now defined by (\ref{e6}) as
$p_{+}H^{-1}$ and $p_{-}H^{-1}$ respectively, are operators, so that the
expressions for $g_{ij}$ do not split cleanly into ``classical'' and
``quantum'' factors.

Since$H^{2}=p^{2}=p_{+}^{2}+p_{-}^{2}$ we have $\exp (aH)=\cosh
(ap)+(H/p)\sinh (ap)$. If we use this result in (\ref{e5a}) we find%
\[
\beta _{\pm }=[(\hslash /2p^{2})B_{\pm }\sin (\frac{2\Omega p}{\hslash }%
)+p_{\pm }\Omega /p](H/p)+(i\hslash /2p^{2})B_{\pm }\cos (\frac{2\Omega p}{%
\hslash })+C_{\pm }
\]%
The metric components can then be written as
\begin{equation}
g_{ii}=(\exp D_{i})R^{2}\exp [\lambda _{i}(t)+\gamma _{i}(t)H/p]  \label{e10}
\end{equation}%
\quad \quad \quad \quad \quad where

\quad
\begin{eqnarray}
\lambda _{\pm } &=&-2\Omega +i[\hslash (B_{+}\pm \sqrt{ 3}B_{-})/p^{2}]\cos (%
\frac{2\Omega p}{\hslash })  \nonumber \\
\lambda _{3} &=&-2\Omega -2i(\hslash B_{+}/p^{2})\cos (\frac{2\Omega p}{%
\hslash })  \label{e11} \\
\gamma _{\pm } &=&(\hslash /p^{2})(B_{+}\pm \sqrt{ 3}B_{-})\sin (\frac{%
2\Omega p}{\hslash })+2(p_{+}\pm \sqrt{ 3}p_{-})\Omega /p  \nonumber \\
\gamma _{3} &=&-2[(\hslash /p^{2})B_{+}\sin (\frac{2\Omega p}{\hslash }%
)+2p_{+}\Omega /p]  \nonumber
\end{eqnarray}%
with $D_{1}=2(C_{+}+\surd 3C_{-})$ , $D_{2}=2(C_{+}-\surd 3C_{-})$ , and $%
D_{3}=-4C_{+}$ . (We also have dropped the summation convention.)

For brevity we put $\phi _{i}=\lambda _{i}+\gamma _{i}H/p$ so that $%
g_{ii}=\exp \left( D_i\right) R^{2}\exp \left( \phi_i\right) $. The
contravariant components $g^{ii}=R^{-2}\exp \left( -\phi _{i}\right) \exp
\left( -D_{i}\right) $ will then be the correct inverse of $g_{ii}$ .

The temporal evolution of the metric is now rather complicated. The quantum
feature par excellence is the fact that $B$ and $H$ are operators which can
be represented by $2\times 2$ matrices in the present model. But we should
also note that the simple complex exponential or circular functions which
describe the Zitterbewegung of a particle are now themselves the arguments
of the exponential functions which give the metric in (\ref{e8}) and (\ref%
{e9}). As $t\rightarrow 0$ and $\Omega \rightarrow \infty $, the magnitude
of the metric components will oscillate an infinite number of times.

\section{Quantum Cosmology in the Heisenberg Picture: The Curvature}

In order to investigate the behavior of space-time near $t=0$ ($\Omega
=\infty $), the true singularity of the classical model, we must of course
examine not simply the metric but the curvature.

In the classical case we can use the cosmic time $t$ and write the metric
corresponding to (\ref{e9}) as $g_{00}=-1$, $g_{aa}=t^{2s_{a}}$ with
appropriate scaling. The non-vanishing components of the Riemann tensor are
given by $R_{a0a0}=-\Gamma _{0aa,0}+\Gamma _{aa0}\Gamma
_{0aa}=-s_{a}(s_{a}-1)t^{2s_{a}-2}$ and the Kretschmann scalar $K=R_{\alpha
\beta \gamma \delta }R^{\alpha \beta \gamma \delta }$
\begin{equation}
K=-16\frac{s_{1}s_{2}s_{3}}{t^{4}}  \label{e11a}
\end{equation}%
which diverges as $t\rightarrow 0$.

The quantum theory is more complicated and there is considerable ambiguity
because of the non-commutativity of operators. There are three ways in which
this complicates matters. First, there is the factor-ordering problem, which
always arises when a transition is made from a classical to a quantum
theory. Secondly, the operator scale factors in the metric which we noted
above must be taken into account. Finally, we have to be aware that the $B$%
's must have an algebraic structure ensuring that $\sigma $'s (which now in
the Heisenberg picture are operators) have the correct algebra.

If we choose the usual order of factors for the Christoffel symbols we
obtain
\[
\Gamma _{m0n}=\frac{1}{2}\delta _{mn}\phi _{m}^{\prime },\qquad 
_{0mn}=\frac{1}{2}R_{0}^{2}(\exp D_{m})\delta _{mn}\phi _{m}^{\prime
}(\exp \phi _{m}),
\]%
where a prime denotes the time derivative. We then calculate curvature
components with the ordering of factors given by
\begin{equation}
\quad R_{\mu \nu \kappa }^{\lambda }=\Gamma _{\mu \nu ,\kappa }^{\lambda
}-\Gamma _{\mu \kappa ,\nu }^{\lambda }+\Gamma _{\mu \nu }^{\sigma }\Gamma
_{\kappa \sigma }^{\lambda }-\Gamma _{\mu \kappa }^{\sigma }\Gamma _{\nu
\sigma }^{\lambda }  \label{e12a}
\end{equation}
It would be desirable, as we will see below, for factors to be ordered in
such a way that all occurrences of $\exp D_{m}$ in the curvature occur to
the left of other factors. Unfortunately this would require a different
ordering of factors for different components, so that such a result would
not be covariant. Instead of trying to indicate all the possibilities we
will simply follow the prescription in (\ref{e12a}) and write the results as
follows.

\begin{eqnarray}
R_{\ 0n0}^{m} &=&\frac{1}{2}\delta _{mn}\left( \phi _{m}^{\prime
\prime }+\phi _{m}^{\prime 2}\right)  \nonumber \\
R_{\ mn0}^{0} &=&\frac{1}{2}\exp D_{m}(R_{0}^{2})\delta _{mn}\left[
\phi _{m}^{\prime \prime }+\frac{1}{2}\phi _{m}^{\prime 2}\right]
\exp \phi
_{m}+C.T  \label{e13} \\
R_{\ mnk}^{l} &=&\frac{1}{2}\exp D_{m}(R_{0}^{2})\left( \delta
_{mn}\delta _{kl}-\delta _{mk}\delta _{nl}\right) \phi _{m}^{\prime
}\phi _{l}^{\prime }+C.T  \nonumber
\end{eqnarray}%
Here C.T. denotes ``commutator terms,'' distinctively quantum mechanical
expressions involving $\hslash $, which result from switching the order of
factors containing $D$'s (and thus $C$'s) and $p$'s.

We then find that the mixed Ricci tensor components are

\begin{eqnarray}
R_{0}^{0} &=&\left( \frac{2}{3p^{2}t^{2}}\right) \left\{ B^{2}\left[ \cos
^{2}\left( \frac{2\Omega p}{\hslash }\right) -\sin ^{2}\left( \frac{2\Omega p%
}{\hslash }\right) \right] \right.  \nonumber \\
&&\left. +2{\bf p} {\bf B}\cos \left( \frac{2\Omega p}{\hslash }\right)
+2i\sin \left( \frac{2\Omega p}{\hslash }\right) \left[ B^{2}\cos \left(
\frac{2\Omega p}{\hslash }\right) +{\bf p} {\bf B}\right] \frac{H}{p}%
\right\} +\rm{C.T.}  \nonumber \\
R_{1}^{1} &=&-2i\left( \frac{B_{+}+\sqrt{ 3}B_{-}}{9\hslash t^{2}}\right) %
\left[ \cos \left( \frac{2\Omega p}{\hslash }\right) +i\sin \left( \frac{%
2\Omega p}{\hslash }\right) \frac{H}{p}\right] +\rm{C.T}  \label{e14} \\
R_{2}^{2} &=&-2i\left( \frac{B_{+}-\sqrt{ 3}B_{-}}{9\hslash t^{2}}\right) %
\left[ \cos \left( \frac{2\Omega p}{\hslash }\right) +i\sin \left( \frac{%
2\Omega p}{\hslash }\right) \frac{H}{p}\right] +\rm{C.T}  \nonumber \\
R_{3}^{3} &=&4i\left( \frac{B_{+}}{9\hslash t^{2}}\right) \left[ \cos \left(
\frac{2\Omega p}{\hslash }\right) +i\sin \left( \frac{2\Omega p}{\hslash }%
\right) \frac{H}{p}\right] +\rm{C.T}  \nonumber
\end{eqnarray}

Because of cancellations the Ricci scalar $R$ is simply equal to $R_{0}^{0}$%
. We have written\thinspace $\ {\bf p} {\bf B=}p_{+}B_{+}+p_{-}B_{-}$ and $%
B^{2}=$ $B_{+}^{2}+B_{-}^{2}.$ In the calculation of the mixed Ricci
components, all occurrences of the commutator of $B_{+}$ and $B_{-}$ cancel
out.

Calculation of the Kretschmann scalar is now considerably more complicated,
even if the problems of factor ordering are ignored. In addition, the need
for four factors of metric components to lower and raise indices introduces
a great many more choices for their ordering, and the non-commutativity of
the Bs can no longer be ignored. The full expression for K with the choices
we have made for factor ordering is quite lengthy. It can be written as
\begin{equation}
K=K_{0}+K_{1}\frac{H}{p}  \label{e15}
\end{equation}%
where%
\begin{eqnarray}
K_{0} &=&\frac{4}{27\hslash ^{2}t^{4}}\left[ 8B^{2}\left( 1-2\cos ^{2}\left(
\frac{2\Omega }{\hslash }p\right) \right) +5\hslash ^{2}\right] +\frac{%
32\sin \left( \frac{2\Omega }{\hslash }p\right) }{27\hslash t^{4}p}\left[
2B^{2}\cos \left( \frac{2\Omega }{\hslash }p\right) +\vec{p}\cdot \vec{B}%
\right]   \nonumber \\
&&-\frac{8}{27\hslash t^{4}p^{2}}\left[ -3\hslash \left( B^{2}-p^{2}\right)
+8i\left( B_{+}^{2}p_{+}-B_{-}^{2}p_{+}-2p_{-}B_{+}B_{-}\right) \right]
\nonumber \\
&&+\frac{16i}{27\hslash t^{4}p^{2}}\cos \left( \frac{2\Omega }{\hslash }%
p\right) \left\{ 3i\hslash \vec{p}\cdot \vec{B}-6B_{+}\left( B_{+}+\sqrt{3}%
B_{-}\right) \left( B_{+}-\sqrt{3}B_{-}\right) +2\left(
p_{+}^{2}B_{+}-p_{-}^{2}B_{+}-2B_{-}p_{+}p_{-}\right) \right\}   \nonumber \\
&&+\frac{16i\cos ^{2}\left( \frac{2\Omega }{\hslash }p\right) }{27\hslash
t^{4}p^{2}}\left[ 3i\hslash B^{2}+8\left(
B_{+}^{2}p_{+}-B_{-}^{2}p_{+}-2p_{-}B_{+}B_{-}\right) \right]   \nonumber \\
&&-\frac{128i\cos ^{3}\left( \frac{2\Omega }{\hslash }p\right) }{27\hslash
t^{4}p^{2}}B_{+}\left( B_{+}+\sqrt{3}B_{-}\right) \left( B_{+}-\sqrt{3}%
B_{-}\right)   \nonumber \\
&&-\frac{128i\sin \left( \frac{2\Omega }{\hslash }p\right) \cos ^{2}\left(
\frac{2\Omega }{\hslash }p\right) }{27t^{4}p^{3}}\left( B_{+}\left( B_{+}+%
\sqrt{3}B_{-}\right) \left( B_{+}-\sqrt{3}B_{-}\right) \right)   \nonumber \\
&&-\frac{64i\sin \left( \frac{2\Omega }{\hslash }p\right) }{9t^{4}p^{3}}\cos
\left( \frac{2\Omega }{\hslash }p\right) \left(
B_{+}^{2}p_{+}-B_{-}^{2}p_{+}-2p_{-}B_{+}B_{-}\right)   \nonumber \\
&&-\frac{32i\sin \left( \frac{2\Omega }{\hslash }p\right) }{27t^{4}p^{3}}%
\left(
-6B_{-}p_{+}p_{-}+3B_{+}B_{-}^{2}+3B_{+}p_{+}^{2}-3B_{+}p_{-}^{2}-B_{+}^{3}%
\right)   \nonumber \\
&&+\frac{32}{3t^{4}p^{4}}B^{4}\cos ^{4}\left( \frac{2\Omega }{\hslash }%
p\right) +\frac{64}{3t^{4}p^{4}}\cos ^{3}\left( \frac{2\Omega }{\hslash }%
p\right)
(B_{-}^{3}p_{-}+B_{+}B_{-}^{2}p_{+}+B_{+}^{3}p_{+}+B_{+}^{2}B_{-}p_{-})
\nonumber \\
&&+\frac{16}{3t^{4}p^{4}}\cos ^{2}\left( \frac{2\Omega }{\hslash }p\right)
(B_{+}^{2}p_{-}^{2}+3B_{-}^{2}p_{-}^{2}-2B^{4}+3B_{+}^{2}p_{+}^{2}+4B_{-}p_{-}B_{+}p_{+}-4B_{+}^{2}B_{-}^{2}+B_{-}^{2}p_{+}^{2})
\nonumber \\
&&+\frac{16}{3t^{4}p^{4}}\cos \left( \frac{2\Omega }{\hslash }p\right)
(p_{-}^{3}B_{-}+p_{+}^{3}B_{+}+p_{+}p_{-}^{2}B_{+}-3B_{+}^{2}B_{-}p_{-}-3B_{+}^{3}p_{+}-3B_{-}^{3}p_{-}-3B_{+}B_{-}^{2}p_{+}+B_{-}p_{+}^{2}p_{-})
\nonumber \\
&&+\frac{4}{3t^{4}p^{4}}%
(p^{4}+B^{4}-2B_{+}^{2}p_{-}^{2}-6B_{+}^{2}p_{+}^{2}-6B_{-}^{2}p_{-}^{2}-8B_{-}p_{-}B_{+}p_{+}-2B_{-}^{2}p_{+}^{2})
\label{e15a}
\end{eqnarray}%
and%
\begin{eqnarray}
K_{1} &=&-\frac{64i}{27t^{4}\hslash ^{2}}\sin \left( \frac{2\Omega }{\hslash
}p\right) \cos \left( \frac{2\Omega }{\hslash }p\right) B^{2}+\frac{32i}{%
27t^{4}\hslash p}\left( B^{2}\left[ 1-2\cos ^{2}\left( \frac{2\Omega }{%
\hslash }p\right) \right] -2\vec{p}\cdot \vec{B}\cos \left( \frac{2\Omega }{%
\hslash }p\right) \right)   \nonumber \\
&&-\frac{128\sin \left( \frac{2\Omega }{\hslash }p\right) \cos ^{2}\left(
\frac{2\Omega }{\hslash }p\right) }{27\hslash t^{4}p^{2}}\left( B_{+}\left(
B_{+}+\sqrt{3}B_{-}\right) \left( B_{+}-\sqrt{3}B_{-}\right) \right)
\nonumber \\
&&-\frac{16\sin \left( \frac{2\Omega }{\hslash }p\right) \cos \left( \frac{%
2\Omega }{\hslash }p\right) }{27\hslash t^{4}p^{2}}(3i\hslash B^{2}+8\left(
B_{+}^{2}p_{+}-B_{-}^{2}p_{+}-2p_{-}B_{+}B_{-}\right) )  \nonumber \\
&&-\frac{16\sin \left( \frac{2\Omega }{\hslash }p\right) }{27\hslash
t^{4}p^{2}}\left( 3i\hslash \vec{p}\cdot \vec{B}+2\left(
p_{+}^{2}B_{+}-p_{-}^{2}B_{+}-2p_{-}p_{+}B_{-}\right)
+6B_{+}B_{-}^{2}-2B_{+}^{3}\right)   \nonumber \\
&&-\frac{128\cos ^{3}\left( \frac{2\Omega }{\hslash }p\right) }{27t^{4}p^{3}}%
\left( B_{+}\left( B_{+}+\sqrt{3}B_{-}\right) \left( B_{+}-\sqrt{3}%
B_{-}\right) \right) -\frac{64\cos ^{2}\left( \frac{2\Omega }{\hslash }%
p\right) }{9t^{4}p^{3}}\left(
B_{+}^{2}p_{+}-B_{-}^{2}p_{+}-2p_{-}B_{+}B_{-}\right)   \nonumber \\
&&-\frac{32\cos \left( \frac{2\Omega }{\hslash }p\right) }{9t^{4}p^{3}}%
\left( 3B_{+}B_{-}^{2}-B_{+}^{3}+\left(
p_{+}^{2}B_{+}-p_{-}^{2}B_{+}-2p_{-}p_{+}B_{-}\right) \right)   \nonumber \\
&&-\frac{32}{27t^{4}p^{3}}%
(6B_{+}B_{-}p_{-}+3B_{-}^{2}p_{+}+p_{+}^{3}-3B_{+}^{2}p_{+}-3p_{-}^{2}p_{+})
\label{e15b} \\
&&+\frac{32i\sin \left( \frac{2\Omega }{\hslash }p\right) \cos ^{3}\left(
\frac{2\Omega }{\hslash }p\right) }{3t^{4}p^{2}}B^{4}+\frac{64i\sin \left(
\frac{2\Omega }{\hslash }p\right) \cos ^{2}\left( \frac{2\Omega }{\hslash }%
p\right) }{3t^{4}p^{2}}%
(p_{+}B_{+}B_{-}^{2}+B_{-}^{3}p_{-}+B_{+}^{3}p_{+}+3B_{+}^{2}B_{-}p_{-})
\nonumber \\
&&+\frac{16i\sin \left( \frac{2\Omega }{\hslash }p\right) \cos \left( \frac{%
2\Omega }{\hslash }p\right) }{3t^{4}p^{2}}\left(
-2B_{+}^{2}B_{-}^{2}+4B_{+}B_{-}p_{-}p_{+}+3p_{-}^{2}B_{-}^{2}+3p_{+}^{2}B_{+}^{2}+B_{+}^{2}p_{-}^{2}-B_{-}^{4}+B_{-}^{2}p_{+}^{2}-B_{+}^{4}\right)
\nonumber \\
&&+\frac{16i\sin \left( \frac{2\Omega }{\hslash }p\right) }{3t^{4}p^{2}}%
\left(
-p_{+}B_{+}^{3}+p_{-}^{2}p_{+}B_{+}+p_{+}^{3}B_{+}+p_{+}^{2}p_{-}B_{+}-B_{+}B_{-}^{2}p_{+}-B_{+}^{2}B_{-}p_{-}+B_{-}p_{-}^{3}-B_{-}^{3}p_{-}\right)
\nonumber
\end{eqnarray}%
The terms $\tilde{K}$ that do not involve any distinctively quantum factors,
and that are not affected by factor ordering complications, can be written as%
\begin{equation}
\tilde{K}=\frac{8}{27t^{4}}\left\{ 10+\left[ 4\left( \frac{%
3p_{-}^{2}-p_{+}^{2}}{p^{3}}\right) \right] p_{+}\frac{H}{p}\right\}
\label{e16}
\end{equation}%
This is identical with the classical result given by (\ref{e11a}). This
means that while the magnitude of this curvature invariant will undergo
oscillations of increasing rapidity as $t\rightarrow 0$, there will be an
overall unbounded increase beyond any bounds according to the same $t^{-4}$
law that obtains in the classical case. The leading divergence of the
curvature is not removed by quantization.

There are, of course, profound differences between classical and quantum
curvatures in other regards. As is well known, the special case of the
Kasner metric with $s_{1}=s_{2}=0$ and $s_{3}=1$ in the classical theory is
simply flat space-time. This is no longer so in quantum theory because of
the additional exponential factors in the expressions (\ref{e9}) for the
metric.

Use of the Heisenberg picture thus gives some novel insight into the quantum
dynamics of the simplest model universe of the Bianchi type, and shows in
particular that the initial singularity is not avoided by quantization.
There are no questions about boundary conditions on a wave function or the
interpretation of it to cloud this result. Unfortunately the procedure
followed here cannot be extended immediately to more complicated models.
Just as straightforward integration of the equations of motion for the Dirac
electron works only for a free particle and not when an external potential
is present, the curvature term which would be added to right side of (\ref%
{e1}) frustrates attempts to follow the same procedure for spaces of other
Bianchi types.

In a sense our result is not too surprising. When we solve the Dirac eqn for
a particle we recover the classical motion of the particle together with the
distinctive Zitterbewegung effects. In the cosmological case a similar thing
happens -- we recover the classical cosmological motion corrected by quantum
effects. But in this case the classical solution is divergent, and the
question of interest is whether or not the quantum effects can cancel this
divergence. In particular, does the expectation value of the Kretschmann
scalar blow up as $t\rightarrow 0$ in the quantum case as it does in the
classical one?

We have not been able to answer this question definitively because of the
huge number of possible factor orderings in $K$ that arise in different
ways. The most straightforward way of writing $K$ does give a divergent
result, and this may suggest that hopes for an elimination of the initial
cosmological singularity by quantum theory will be disappointed. But it is
not impossible that some choice of factor ordering will eliminate this
divergence.

Use of the Heisenberg picture thus gives some novel insight into the quantum
dynamics of the simplest model universe of the Bianchi type, and shows in
particular that the initial singularity is not avoided by quantization.
There are no questions about boundary conditions on a wave function or the
interpretation of it to cloud this result. Unfortunately the procedure
followed here cannot be extended immediately to more complicated models.
Just as straightforward integration of the equations of motion for the Dirac
electron works only for a free particle and not when an external potential
is present, a term involving the 3-space curvature term which would be added
to right side of (\ref{e1}) frustrates attempts to follow the same procedure
for spaces of other Bianchi types. The problem would, of course, be even
more difficult for nonhomogeneous universes or other space-times\cite{ref5}%
.\bigskip

{\Large Acknowledgments}

This work was supported in part by the Natural Sciences and Engineering
Research Council of Canada.\


\medskip

\begin{thebibliography}{9}
\bibitem{ref1} B. DeWitt, Phys. Rev. {\bf 160}, 1113 (1967): C. Misner,
``Classical and Quantum Dynamics of a Closed Universe'' M. Carmeli, S.I.
Fickler and L. Witten, {\it Relativity} (Plenum, New York, 1970), pp.55-79:
M. Ryan, {\it Hamiltonian Cosmology} (Springer-Verlag, New York, 1972). The
notation of the latter monograph is used in this paper.

\bibitem{ref2} G.L. Murphy, Found. Phys. {\bf 4,} 351, (1984).

\bibitem{ref3} E. Schr\"{o}dinger, Sitzungb. d. Berlin. Akad. {\bf XXIV},
418 (1930): P.A.M. Dirac, {\it The Principles of Quantum Mechanics}, 4th ed.
(Clarendon, Oxford, 1958), pp.261-263.

\bibitem{ref4} E. Kasner, Am. J. Math. {\bf 43}, 217 (1921).

\bibitem{ref5} M. Ryan, {\it Hamiltonian Cosmology} (Springer-Verlag, New
York, 1972) p 25.
\end{thebibliography}
\end{document}